\documentclass[10pt]{article}
\usepackage[centertags]{amsmath}
\usepackage{amsfonts}
\usepackage{amssymb}
\usepackage{amsthm}
\usepackage{epsfig}
\usepackage{color}
\usepackage{array}
\usepackage{mathrsfs}
\allowdisplaybreaks[4]
\usepackage{indentfirst}
\usepackage{amssymb,color}
\allowdisplaybreaks
\definecolor{c20}{rgb}{1.00,0.00,0.00}
\definecolor{c30}{rgb}{0.,0.,1.}
\definecolor{c40}{rgb}{1,0.1,0.7}
\definecolor{c50}{rgb}{1,0,0}
\definecolor{c60}{rgb}{1,0.9,0.1}
\definecolor{c70}{rgb}{0.50,1.00,0.00}
\definecolor{c80}{rgb}{0.00,1.00,0.00}

\def\P{\operatorname*{\mathbb{P}}}

\def\E{\operatorname*{\mathrm{E}}}

\def\MSE{\operatorname*{\mathrm{MSE}}}

\date{}

\newtheorem{theorem}{Theorem}[]

\numberwithin{equation}{section}

\def\P{\operatorname*{\mathbb{P}}}
\def\R{\operatorname*{\mathbb{R}}}

\def\E{\operatorname*{\mathbf{E}}}

\def\M{\operatorname*{\mathbf{M}}}
\begin{document}
\title{Asymptotics and statistical inferences on independent and non-identically distributed bivariate Gaussian triangular arrays}
\author{{  Xin Liao\quad Zuoxiang Peng\thanks{Corresponding author. Email: pzx@swu.edu.cn}}\\
{\small\it School of Mathematics and Statistics, Southwest
University, 400715 Chongqing, China }}
\maketitle

\begin{quote}
{\bf Abstract}~~In this paper, we establish the first and the
second-order asymptotics of distributions of normalized maxima of
independent and non-identically distributed bivariate Gaussian
triangular arrays, where each vector of the $n$th row follows from a
bivariate Gaussian distribution with correlation coefficient being a
monotone continuous positive function of $i/n$. Furthermore,
parametric inference for this unknown function is studied. Some
simulation study and real data sets analysis are also presented.



{\bf Running Title}~~Asymptotics and statistical inferences on
bivariate Gaussian triangular arrays

\end{quote}

\section{Introduction}
\label{sec1}

Let $\{(X_{ni},Y_{ni}),1\leq i \leq n, n\geq 1 \}$ be independent
bivariate Gaussian triangular arrays, and let $\rho_{ni}$ denote the
correlation coefficient of $(X_{ni},Y_{ni})$, $1\le i\le n$. The
bivariate maxima $\M_{n}$ are defined componentwise by
\[{\M}_{n}=(M_{n1},M_{n2})
=\left(\max_{1 \leq i\leq n} X_{ni}, \max_{1\leq i \leq n}
Y_{ni}\right). \] For the asymptotic distribution of $\M_{n}$,
Sibuya (1960) showed that $M_{n1}$ and $M_{n2}$ are asymptotic
independent if $\rho_{ni}=\rho\in (-1,1)$, which coincides with the
tail asymptotic independence of Gaussian copula, see Embrechts et
al. (2002). For the case of $\rho_{ni}=\rho_{n}$, H\"{u}sler and
Reiss (1989) derived that
\begin{eqnarray}\label{eq1.1}
& & \lim_{n\to\infty}\sup_{x\in \R, y\in
\R}\left|\P\left({M}_{n1}\leq b_{n}+\frac{x}{b_{n}}, {M}_{n2}\leq
b_{n}+\frac{y}{b_{n}} \right)- H_{\lambda}(x,y)\right|=0
\end{eqnarray}
provided that the following H\"{u}sler-Reiss condition
\begin{equation}\label{eq1.2}
\lim_{n\to \infty}b_{n}^{2}(1-\rho_{n})=2\lambda^{2}
\end{equation} holds with $ \lambda \in [0,\infty]$ ( the converse assertion is proved by Kabluchko et al. (2009)),
where the norming constant $b_{n}$ satisfies
\begin{equation}\label{eq1.3}
1-\Phi(b_{n})=\frac{1}{n},
\end{equation}
where $H_{\lambda}(x,y)$, the so-called H\"{u}sler-Reiss max-stable
distribution, is given by
\[H_{\lambda}(x,y)=\exp\left(
-\Phi\left(\lambda+\frac{x-y}{2\lambda}\right)e^{-y} -
\Phi\left(\lambda+\frac{y-x}{2\lambda}\right)e^{-x}  \right)\] with
$\Phi(x)$  standing for the standard Gaussian distribution.
Obviously, components of $\M_{n}$ are asymptotic dependent when
$\lambda<\infty$.

For H\"{u}sler-Reiss model, the asymptotic behavior of the dynamic
copula version of normalized $\M_{n}$ has also been studied in
recent literature. Under the H\"{u}sler-Reiss condition
\eqref{eq1.2}, Frick and Reiss (2013) considered the asymptotic
behaviors of the distribution of $(n(\max_{1\leq i \leq
n}\Phi(X_{ni})-1), n(\max_{1\leq i \leq n}\Phi(Y_{ni})-1))$.
Allowing $\rho_{ni}$ to depend on both $i$ and $n$, Liao et al.
(2014a) extended the result in Frick and Reiss (2013) by assuming
that
\begin{equation}\label{eq2.1}
\rho_{ni}=1-m(i/n)/\log n
\end{equation}
for some positive function $m(x)$. For other work related to
H\"{u}sler-Reiss model and its extensions, see, e.g., Hashorva
(2005, 2006), Hashorva et al. (2012), Hashorva and Weng (2013),
Kabluchko (2011), Engelke et al. (2014), Das et al. (2014) and
reference therein.

The objective of this paper is to derive the first and the
second-order distributional expansions of the dynamic
H\"{u}sler-Reiss model with $\rho_{ni}$ given by \eqref{eq2.1} and
establish statistical inferences related to the function $m(x)$. For
the convergence rates and higher-order expansions of univariate
extremes, we refer to de Haan and Resnick (1996), Nair (1981), Liao
et al. (2014b) and reference therein. For the convergence rates of
bivariate extremes, see de Haan and Peng (1997) for the general
case. For the special case of the bivariate H\"{u}sler-Reiss model,
Hashorva et al. (2014) established the higher-order distributional
expansions of $\M_{n}$, and Liao and Peng (2014) established the
uniform convergence rate of \eqref{eq1.1}. Liao and Peng (2015) also
derived the second-order expansion of the joint distribution of
normalized maximum and minimum. So far, there are no studies on the
convergence and distributional expansion of $\M_{n}$ under the
assumption that $(X_{ni},Y_{ni})'s$ are not identically distributed.
The main goal of this paper is to fill this gap. Borrowing the ideas
from Liao et al. (2014a),  we derive in this paper the limit
distribution of the normalized maxima $\M_{n}$ if the function
$m(i/n)$ in \eqref{eq2.1} satisfies some regular conditions, and
establish its second-order distributional expansion provided that
the convergence rate of $\max_{1\le i\le n}m(i/n)$ is given.
Furthermore, parametric estimation of $m(x)$ is considered through
maximum likelihood estimation. The asymptotic properties of the
estimators can be employed to test the condition proposed by
H\"{u}sler and Reiss (1989).

The rest of this paper is organized as follows. In section
\ref{sec2}, we provide the main results and statistical procedures.
A simulation study and some real data analysis are presented in
Section \ref{sec3}. All proofs are given in Section \ref{sec4}.

\section{Methodology}
\label{sec2}

\subsection{Convergence of maxima}

In this section, the limiting distribution and the second-order
expansion of distribution of normalized $\M_{n}$ are provided with
$\rho_{ni}$ satisfying \eqref{eq2.1}. The first result is about the
first-order asymptotic which is stated as follows.
\begin{theorem}\label{th1}
Under the condition \eqref{eq2.1},
\begin{itemize}

\item[(i)] if $\max_{1\leq i \leq n} m(i/n) \to 0$, then for any $x, y\in \R$
\begin{eqnarray*}
\lim_{n\to \infty} \P\left( M_{n1}\leq b_{n} + x/b_{n}, M_{n2}\leq b_{n} + y/b_{n} \right)
= \Lambda(\min(x,y));
\end{eqnarray*}

\item[(ii)] if $\min_{1\leq i \leq n} m(i/n) \to \infty$, then for any $x, y \in \R$
\begin{eqnarray*}
\lim_{n\to \infty} \P\left( M_{n1}\leq b_{n} + x/b_{n}, M_{n2}\leq b_{n} + y/b_{n} \right)
= \Lambda(x)\Lambda(y);
\end{eqnarray*}

\item[(iii)] if $m(s)$ is a continuous positive function on $[0,1]$, then for any $x, y \in \R$
\begin{eqnarray*}
\lim_{n\to \infty} \P\left( M_{n1}\leq b_{n} + x/b_{n}, M_{n2}\leq
b_{n} + y/b_{n} \right) = H(x,y)
\end{eqnarray*}
with
\begin{eqnarray*}
H(x,y)= \exp\left( -e^{-y}\int_{0}^{1} \Phi\left(  \sqrt{m(t)} + \frac{x-y}{2\sqrt{m(t)}}  \right)dt
- e^{-x}\int_{0}^{1}\Phi\left( \sqrt{m(t)} + \frac{y-x}{2\sqrt{m(t)}}  \right)dt \right).
\end{eqnarray*}

\end{itemize}

\end{theorem}

To establish the second-order distributional expansion of normalized
maxima, we consider the following three cases in turn:  $m(t)$ is
monotone and continuous on $[0,1]$;  ${\lim_{n\to
\infty}}\max_{1\leq i \leq n} m(i/n)=0$; and ${\lim_{n\to \infty}}
\min_{1\leq i \leq n} m(i/n)=\infty$.

\begin{theorem}\label{th2}
Under the condition $\eqref{eq2.1}$, assume that $m(t)$ is monotone
and continuous on $[0,1]$, we have
\begin{eqnarray}\label{eq2.2}
& & \lim_{n\to \infty} \frac{\log n}{\log \log n}
\Big( \P\left( M_{n1}\leq b_{n} + x/b_{n}, M_{n2}\leq b_{n} + y/b_{n} \right) - H(x,y) \Big) \nonumber\\
&=& \frac{1}{2}\left(\int_{0}^{1} \sqrt{m(t)}\varphi\left( \sqrt{m(t)}+\frac{y-x}{2\sqrt{m(t)}} \right)dt  \right)
e^{-x}H(x,y),
\end{eqnarray}
where $\varphi(x)$ is the probability density function of standard
Gaussian distribution.
\end{theorem}

\begin{theorem}\label{th3}
Let the norming constant $b_{n}$ be given by \eqref{eq1.3}. Assume
that $\lim_{n\to \infty} (\log n)^{4}\max_{1\leq i \leq n}
m(i/n)=0$, we have
\begin{eqnarray}\label{addeq2.3}
& & \lim_{n\to \infty} (\log n )\Big( \P\left( M_{n1}\leq b_{n} + x/b_{n}, M_{n2}\leq b_{n} + y/b_{n} \right) - \Lambda(\min(x,y))  \Big) \nonumber \\
&=&
\frac{1}{4}\left(\min(x,y))^{2}+2\min(x,y)\right)e^{-\min(x,y)}\Lambda(\min(x,y)).
\end{eqnarray}
\end{theorem}

\begin{theorem}\label{th4}
Let the norming constant $b_{n}$ be given by \eqref{eq1.3}. Assume
that $\lim_{n\to \infty} (\log \log n)/(\min_{1\leq i \leq n}
m(i/n))=0$,
 we have
\begin{eqnarray}\label{addeq2.4}
& & \lim_{n\to \infty} (\log n )\Big( \P\left( M_{n1}\leq b_{n} + x/b_{n}, M_{n2}\leq b_{n} + y/b_{n} \right) - \Lambda(x)\Lambda(y)  \Big) \nonumber \\
&=& \left(\frac{x^{2}+2x}{4}e^{-x} + \frac{y^{2}+2y}{4}e^{-y} \right)\Lambda(x)\Lambda(y).
\end{eqnarray}
\end{theorem}

\subsection{Parametric inference}

Now we consider statistical inference for fitting a parametric form
to the unknown function $m(s)$. Here we consider the family
$m(s)=\alpha + \beta s^{\gamma}$, where $\alpha>0$, $\beta \neq 0$,
$\gamma>0$. Note that when $\beta=0$, $\gamma$ can not be
identified, and when $\gamma=0$, $\alpha$ and $\beta$ cann't be
distinguished, cf. Liao et al. (2014a).

We use the maximum likelihood estimation (MLE) to get the estimator,
which is
\[
\left( \hat{\alpha}, \hat{\beta}, \hat{\gamma} \right) =
\arg\max_{(\alpha,\beta,\gamma)} \left( -n\log 2\pi -
\frac{1}{2}\sum_{i=1}^{n}\log (1-\rho_{ni}^{2}) -
\sum_{i=1}^{n}\frac{X_{ni}^{2}+Y_{ni}^{2}}{2(1-\rho_{ni}^{2})} +
\sum_{i=1}^{n}\frac{\rho_{ni}}{1-\rho_{ni}^{2}}X_{ni}Y_{ni} \right)
.
\]
That is, $\left( \hat{\alpha}, \hat{\beta}, \hat{\gamma} \right)$ is the solution to the following score equations
\begin{equation}\label{eq2.3}
\left\{\begin{array}{ll}
&l_{n1}(\alpha,\beta,\gamma):=\sum_{i=1}^n\left(  \frac{\rho_{ni}}{(\log n)(1-\rho_{ni}^{2})}
+ \frac{(1+\rho_{ni}^{2})X_{ni}Y_{ni}}{(\log n)(1-\rho_{ni}^{2})^{2}}
- \frac{\rho_{ni}(X_{ni}^{2}+Y_{ni}^{2})}{(\log n)(1-\rho_{ni}^{2})^{2}} \right)=0,\\
&l_{n2}(\alpha,\beta,\gamma):=\sum_{i=1}^n\left(  \frac{\rho_{ni}}{(\log n)(1-\rho_{ni}^{2})}
+ \frac{(1+\rho_{ni}^{2})X_{ni}Y_{ni}}{(\log n)(1-\rho_{ni}^{2})^{2}}
- \frac{\rho_{ni}(X_{ni}^{2}+Y_{ni}^{2})}{(\log n)(1-\rho_{ni}^{2})^{2}} \right)
(\frac{i}{n})^{\gamma}=0,\\
&l_{n3}(\alpha,\beta,\gamma):=\sum_{i=1}^n\left(  \frac{\rho_{ni}}{(\log n)(1-\rho_{ni}^{2})}
+ \frac{(1+\rho_{ni}^{2})X_{ni}Y_{ni}}{(\log n)(1-\rho_{ni}^{2})^{2}}
- \frac{\rho_{ni}(X_{ni}^{2}+Y_{ni}^{2})}{(\log n)(1-\rho_{ni}^{2})^{2}} \right)
(\frac{i}{n})^{\gamma}\log(\frac{i}{n})=0.
\end{array}\right.\end{equation}

The following theorem gives the asymptotic normality of the proposed estimator.

\begin{theorem}\label{th5}
Assume that \eqref{eq2.1} holds with $m(s)=\alpha + \beta
s^{\gamma}$ for some $\alpha>0$, $\beta\neq 0$, $\gamma>0$. Then we
have
\begin{eqnarray}\label{eq2.4}
\hat{\Delta}\left( \sqrt{n}(\hat{\alpha}-\alpha), \sqrt{n}(\hat{\beta}-\beta), \sqrt{n}(\hat{\gamma}-\gamma) \right)^{T}
\overset{d}{\to} N\left(0, \Sigma \right),
\end{eqnarray}
where the matrices $\hat{\Delta}$ and $\Sigma$ are given by
\begin{eqnarray*}
\hat{\Delta}=\begin{pmatrix}
\int_{0}^{1}\frac{1}{2(\hat{\alpha}+\hat{\beta} t^{\hat{\gamma}})^{2}}dt
&\int_{0}^{1}\frac{t^{\hat{\gamma}}}{2(\hat{\alpha}+\hat{\beta} t^{\hat{\gamma}})^{2}}dt
&\int_{0}^{1}\frac{\hat{\beta} t^{\hat{\gamma}}\log t}{2(\hat{\alpha}+\hat{\beta} t^{\hat{\gamma}})^{2}}dt\\
\int_{0}^{1}\frac{t^{\hat{\gamma}}}{2(\hat{\alpha}+\hat{\beta} t^{\hat{\gamma}})^{2}}dt
&\int_{0}^{1}\frac{t^{2\hat{\gamma}}}{2(\hat{\alpha}+\hat{\beta} t^{\hat{\gamma}})^{2}}dt
&\int_{0}^{1}\frac{{\hat{\beta}}t^{2\hat{\gamma}}\log t}{2(\hat{\alpha}+\hat{\beta} t^{\hat{\gamma}})^{2}}dt\\
\int_{0}^{1}\frac{{t^{\hat{\gamma}}\log
t}}{2(\hat{\alpha}+\hat{\beta} t^{\hat{\gamma}})^{2}}dt
&\int_{0}^{1}\frac{t^{2\hat{\gamma}}\log
t}{2(\hat{\alpha}+\hat{\beta} t^{\hat{\gamma}})^{2}}dt
&\int_{0}^{1}\frac{\hat{\beta}t^{2\hat{\gamma}}(\log
t)^2}{2(\hat{\alpha}+\hat{\beta} t^{\hat{\gamma}})^{2}}dt
\end{pmatrix}.
\end{eqnarray*}
and
\begin{eqnarray}\label{eq2.5}
\Sigma=\begin{pmatrix} \int_{0}^{1}\frac{1}{2(\alpha+\beta
t^{\gamma})^{2}}dt &\int_{0}^{1}\frac{t^{\gamma}}{2(\alpha+\beta
t^{\gamma})^{2}}dt
&\int_{0}^{1}\frac{t^{\gamma}\log t}{2(\alpha+\beta t^{\gamma})^{2}}dt\\
\int_{0}^{1}\frac{t^{\gamma}}{2(\alpha+\beta t^{\gamma})^{2}}dt
&\int_{0}^{1}\frac{t^{2\gamma}}{2(\alpha+\beta t^{\gamma})^{2}}dt
&\int_{0}^{1}\frac{t^{2\gamma}\log t}{2(\alpha+\beta t^{\gamma})^{2}}dt\\
\int_{0}^{1}\frac{t^{\gamma}\log t}{2(\alpha+\beta
t^{\gamma})^{2}}dt &\int_{0}^{1}\frac{t^{2\gamma}\log
t}{2(\alpha+\beta t^{\gamma})^{2}}dt
&\int_{0}^{1}\frac{t^{2\gamma}(\log t)^2}{2(\alpha+\beta
t^{\gamma})^{2}}dt
\end{pmatrix}
\end{eqnarray}
\end{theorem}

Another interesting parametric form is $m(s)=\alpha+ \beta s$ for
some $\alpha>0$, $\beta \in \R$. In this case, when $\beta=0$,
$m(s)$ becomes constant, which means that the observations $(X_{1},
Y_{1}), \cdots, (X_{n}, Y_{n})$ are independent and identically
distributed random vectors.

\begin{theorem}\label{th6}
Suppose \eqref{eq2.1} holds with $m(s)=\alpha + \beta s$ for some $\alpha >0$,
$\beta \neq 0$. Then we have
\begin{eqnarray}
\begin{pmatrix}
\sqrt{n}\left( \frac{1}{2\hat{\beta}}\log \left( 1+ \frac{\hat{\beta}}{\hat{\alpha}}\right)
-\left( \frac{\hat{\beta}\alpha-\hat{\alpha}\beta}
{2\hat{\alpha}\hat{\beta}(\hat{\alpha}+\hat{\beta})}
+ \frac{\beta}{2\hat{\beta}^2}\log \left( 1+ \frac{\hat{\beta}}{\hat{\alpha}}\right) \right)  \right)
\\
\sqrt{n}\left( \frac{1}{2\hat{\beta}}
-\frac{\hat{\alpha}}{2\hat{\beta}^2}\log \left( 1+ \frac{\hat{\beta}}{\hat{\alpha}}\right)
-\left( \frac{\beta}{2\hat{\beta}^2} + \frac{\hat{\alpha}\beta-\hat{\beta}\alpha}{2\hat{\beta}^{2}(\hat{\alpha}+\hat{\beta})}
-\frac{2\hat{\alpha}\beta-\hat{\beta}\alpha}{2\hat{\beta}^3}\log \left( 1+ \frac{\hat{\beta}}{\hat{\alpha}}\right)
 \right)  \right)
\end{pmatrix}
\overset{d}{\to} N\left( 0, \tilde{\Sigma} \right),
\end{eqnarray}
where $\tilde{\Sigma}$ is given by
\begin{eqnarray*}
\tilde{\Sigma}
=\left( \begin{array}{ll}
\frac{1}{2\alpha (\alpha+ \beta)}
& -\frac{1}{2\beta(\alpha+\beta)}+\frac{1}{2\beta^{2}}\log \left( 1+\frac{\beta}{\alpha}\right) \\
-\frac{1}{2\beta(\alpha+\beta)}+\frac{1}{2\beta^{2}}\log \left( 1+\frac{\beta}{\alpha}\right)
& \frac{1}{2\beta^{2}}\left( 1+\frac{\alpha}{\alpha+\beta}
-\frac{2\alpha}{\beta} \log \left( 1+\frac{\beta}{\alpha}\right) \right)
\end{array} \right).
\end{eqnarray*}
\end{theorem}

\section{Simulation and data analysis}
\label{sec3} In this section we examine the finite sample
performance of the proposed estimators by drawing independent
$(X_{n1},Y_{n1}),\cdots,(X_{nn},Y_{nn})$ with $(X_{ni},Y_{ni})$
following the bivariate Gaussian distribution with coefficient
$\rho_{ni}=1-m(i/n)/\log n$. We consider $n=1000, 3000$ or $10000$,
and repeat $1000$ times.

First we consider $m(s)=\alpha$ with $\alpha=1$ or $10$, and
calculate the average and mean squared error for $\hat{\alpha}$. We
can observe from Table \ref{tab1} that i) the averages of
$\hat{\alpha}$ is near by the true value $\alpha$; ii) small mean
squared errors show the robustness of $\hat{\alpha}$. Next the case
of $m(s)=\alpha+\beta s$ is considered. Table \ref{tab2} reports the
averages and mean squared errors for estimator
$(\hat{\alpha},\hat{\beta})$. As $n$ becomes large, the accuracy of
all estimators improve. Finally, we consider the case of
$m(s)=\alpha+\beta s^{\gamma}$ with sample size $n=10000$. The
simulation shows that all estimators are closer to their true values
with small mean squared errors, cf. Table \ref{tab3} for details.

\begin{table}[htbp]
\caption{Estimators for the case of $m(s)=\alpha$} \label{tab1}
\vspace{0.1in}
\begin{center}
\begin{tabular}{ccccccc}
\hline
& $\alpha=1$ &$\alpha=10$ &$\alpha=1$ &$\alpha=10$ &$\alpha=1$ &$\alpha=10$ \\
&$n=1000$&$n=1000$&$n=3000$&$n=3000$ & $n=10000$& $n=10000$\\
\hline
$\E(\hat{\alpha})$&0.9980043&9.992257&1.001875&9.997267&1.00005&9.999436\\
$\MSE(\hat{\alpha})$&0.002109944&0.02555685&0.0006846334&0.01812748&0.0002033414&0.007965965\\
\hline
\end{tabular}
\end{center}
\end{table}

\begin{table}[htbp]
\caption{Estimators for the case of $m(s)=\alpha+\beta s$ with
$\alpha=1$.} \label{tab2} \vspace{0.1in}
\begin{center}
\begin{tabular}{ccccccc}
\hline
& $\beta=1$ &$\beta=0$ &$\beta=1$ &$\beta=0$ &$\beta=1$ &$\beta=0$\\
&$n=1000$&$n=1000$&$n=3000$&$n=3000$&$n=10000$&$n=10000$\\
\hline
$\E(\hat\alpha)$&1.002458&1.002368&1.001585&0.9973226&0.9996816&1.000073 \\
$\MSE(\hat\alpha)$&0.01125451&0.00864667&0.00385037&0.002605236&0.00124363&0.000797927 \\
$\E(\hat\beta)$&0.9978728& -0.000328&0.9967&0.004754433&1.001072&0.000712807 \\
$\MSE(\hat\beta)$&0.04707124&0.02588997&0.01669121&0.007914633&0.00509091&0.002447169 \\
\hline
\end{tabular}
\end{center}
\end{table}

\begin{table}[htbp]
\caption{Estimators for the case of $m(s)=\alpha+\beta s^{\gamma}$
with $\alpha=\beta=1$.}\label{tab3} \vspace{0.1in}
\begin{center}
\begin{tabular}{ccc|cc|cc}
\hline
&$\E(\hat\alpha)$&$\text{MSE}(\hat\alpha)$&$\E(\hat\beta)$&$\text{MSE}(\hat\beta)$&$\E(\hat\gamma)$&$\text{MSE}(\hat\gamma)$ \\
\hline
$\gamma=0.5$&0.9437925&0.05247714&1.058977&0.04504459&0.5092629&0.03055645\\
$\gamma=1$&0.994503&0.004061697&1.009859&0.005172654&1.01609&0.03928909 \\
$\gamma=1.5$&0.9955838&0.001943133&1.003917&0.005124053&1.504647&0.06171575\\
\hline
\end{tabular}
\end{center}
\end{table}

\begin{figure}[!h]
\centering
\includegraphics[width=7cm,height=5cm]{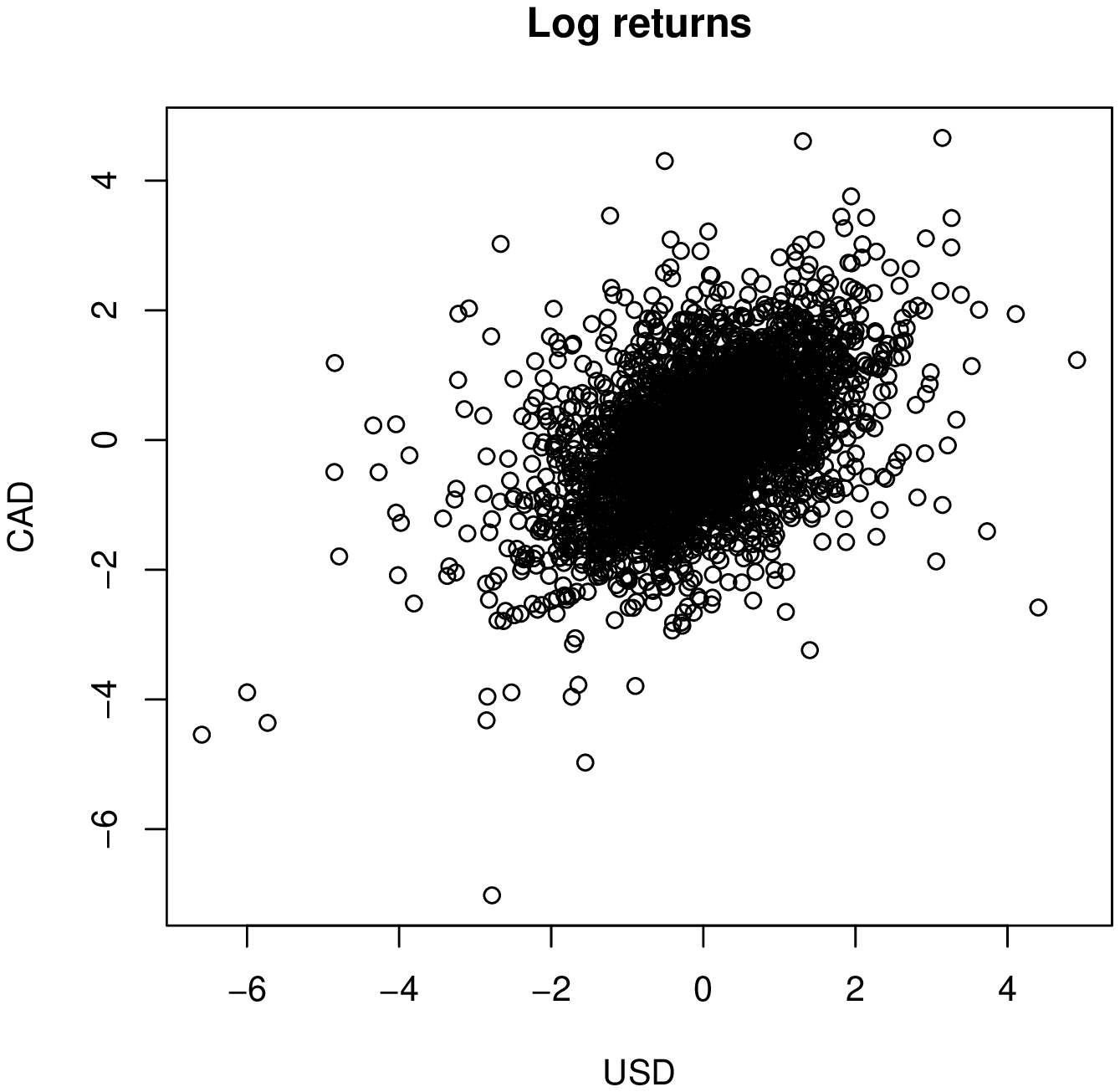}
\includegraphics[width=7cm,height=5cm]{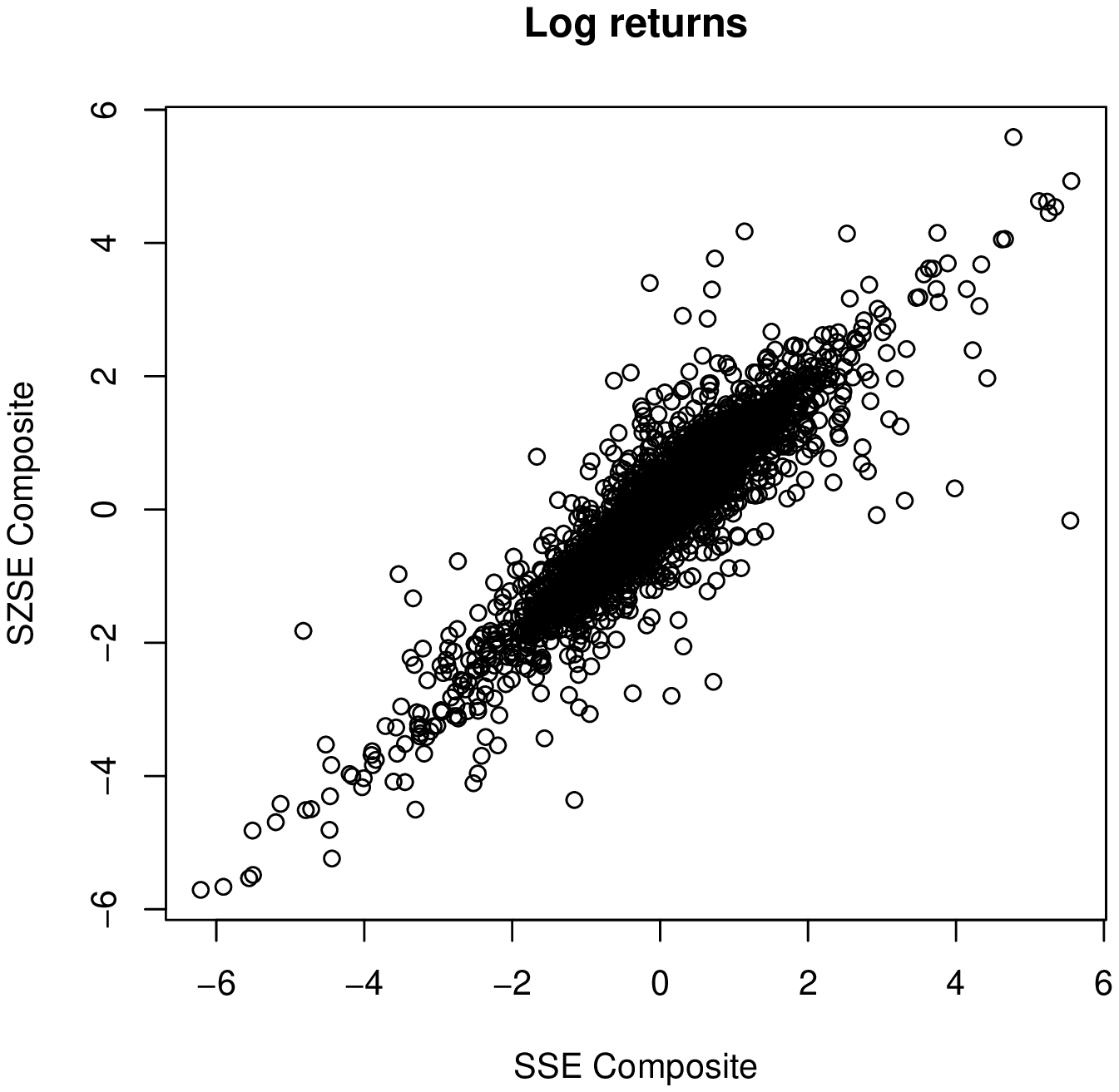}
\includegraphics[width=7cm,height=5cm]{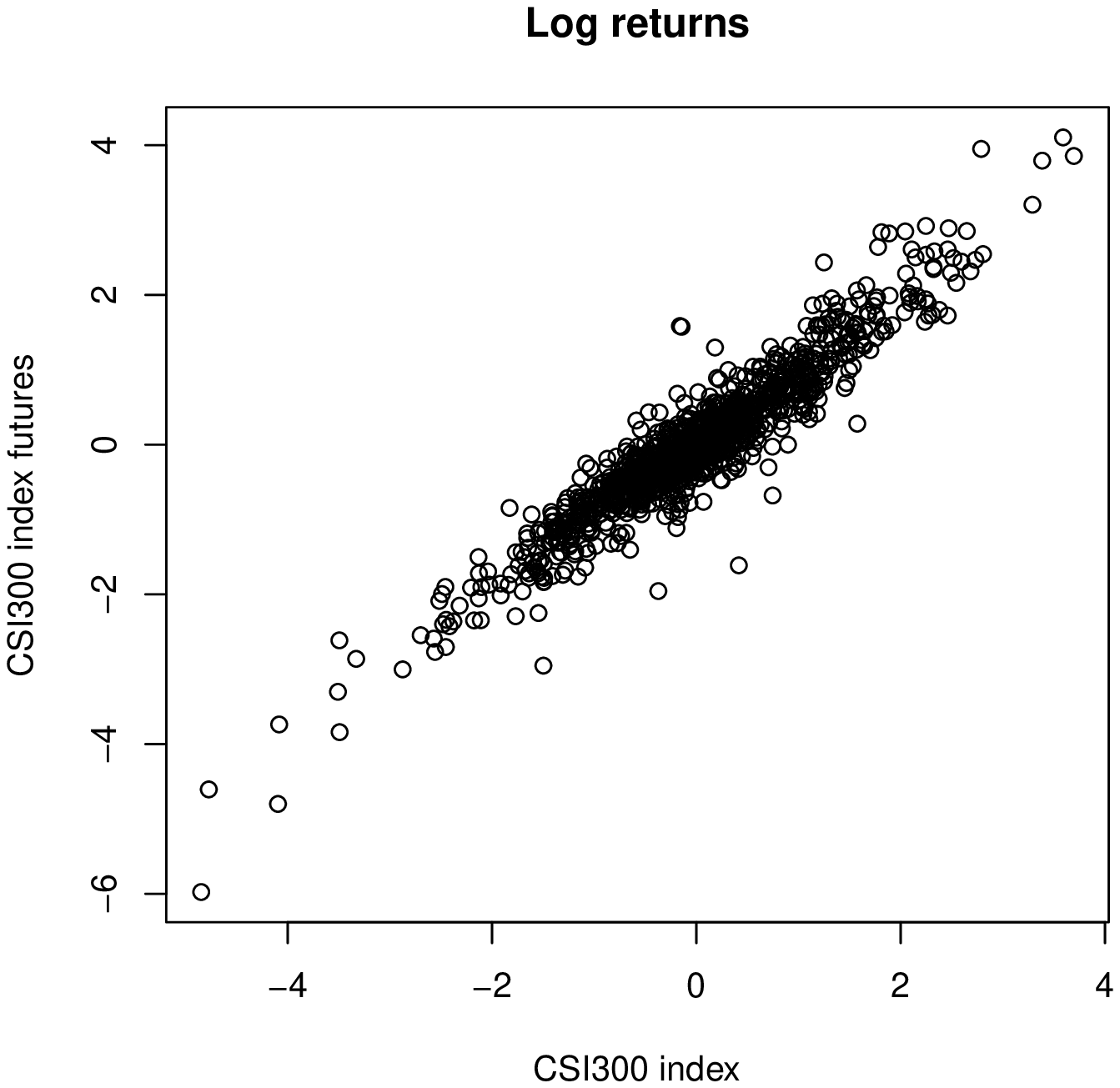}
\includegraphics[width=7cm,height=5cm]{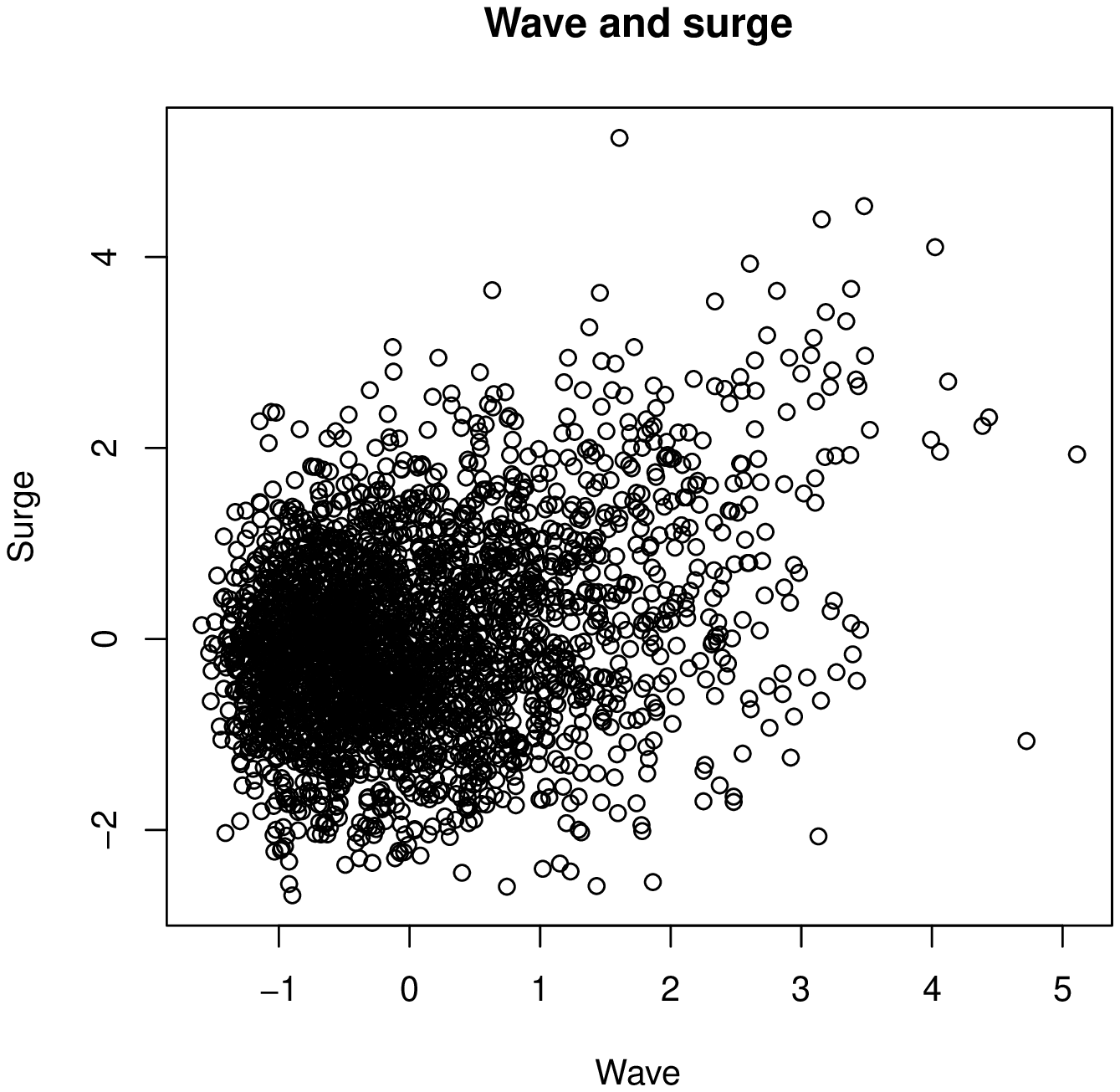}
\caption{log-returns of the exchange rates between US dollar and
British pound and those between Canadian dollar and British pound
(top left); the log-returns of SSE Composite and SZSE Composite from
(top right); the log-returns of the CSI 300 index and CSI 300 index
futures (bottom left); the wave and surge heights in southwest
England (bottom right).} \label{fig1}
\end{figure}

For the applications, we consider four couples of real data sets:
The first
 is the log-returns of the exchange rates between US dollar
and British pound and those between Canadian dollar and British
pound from April 3, 2000 to November 11, 2014. The second is the
log-returns of the Shanghai Stock Exchange composite index (SSE
Composite) and ShenZhen Stock Exchange Composite index (SZSE
Composite) from March 4, 1996 to November 12, 2014. The third is the
log-returns of the CSI 300 index and CSI 300 index futures from
April 16, 2010 to November 13, 2014. The forth is the wave and surge
heights in southwest England which comprise 2894 wave heights and
2894 surge heights. All time series are plotted in Figure
\ref{fig1}.

\begin{figure}[htbp]
\centering
\includegraphics[width=10cm,height=8cm]{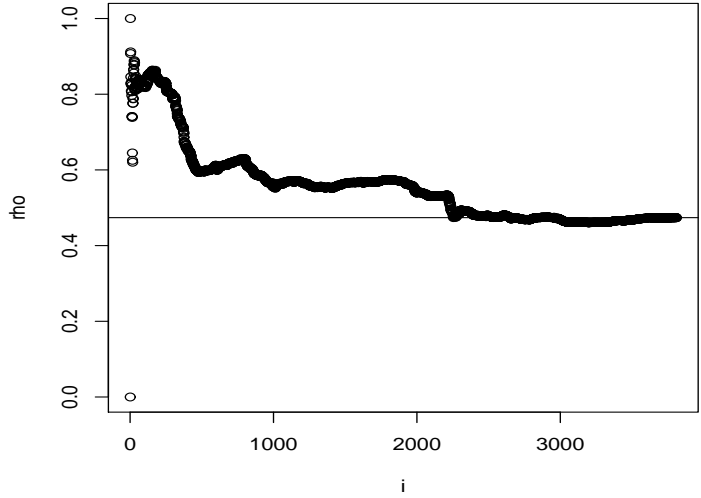}
\caption{Exchange rates. Dotted line represents the sample
correlations, and solid line represents the correlation estimate
$\hat{\rho}=0.4738478$ with $\hat{m}=4.338189$.}
\label{fig2}
\end{figure}

\begin{figure}[htbp]
\centering
\includegraphics[width=10cm,height=8cm]{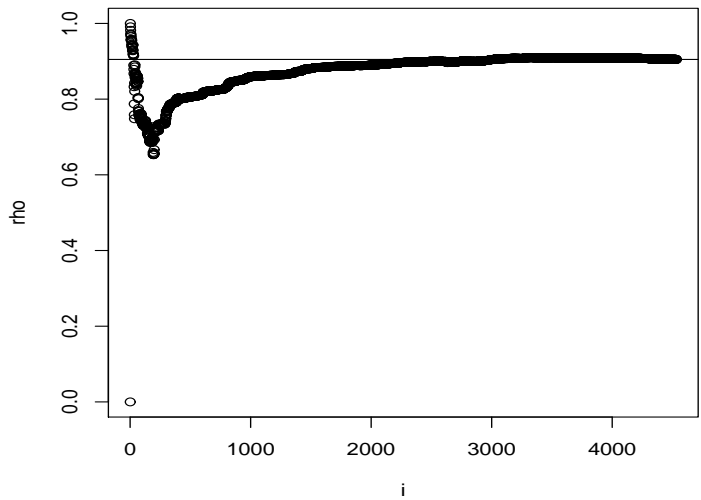}
\caption{SSE Composite and SZSE Composite. Dotted line represents
the sample correlations, and solid line represents the correlation
estimate $\hat{\rho}=0.9048648$ with $\hat{m}=0.8009351$.}
\label{fig3}
\end{figure}

\begin{figure}[htbp]
\centering
\includegraphics[width=10cm,height=8cm]{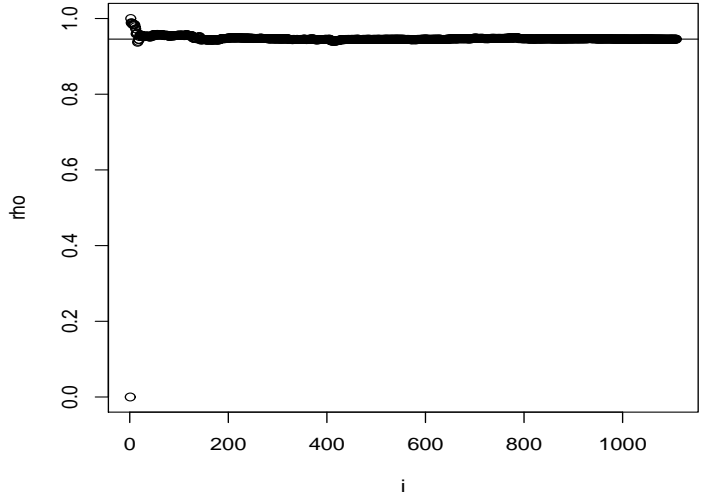}
\caption{CSI 300 index and CSI 300 index futures. Dotted line
represents the sample correlations, and solid line represents the
correlation estimate $\hat{\rho}=0.9455578$ with
$\hat{m}=0.3817058$.}
\label{fig4}
\end{figure}

\begin{figure}[htbp]
\centering
\includegraphics[width=10cm,height=8cm]{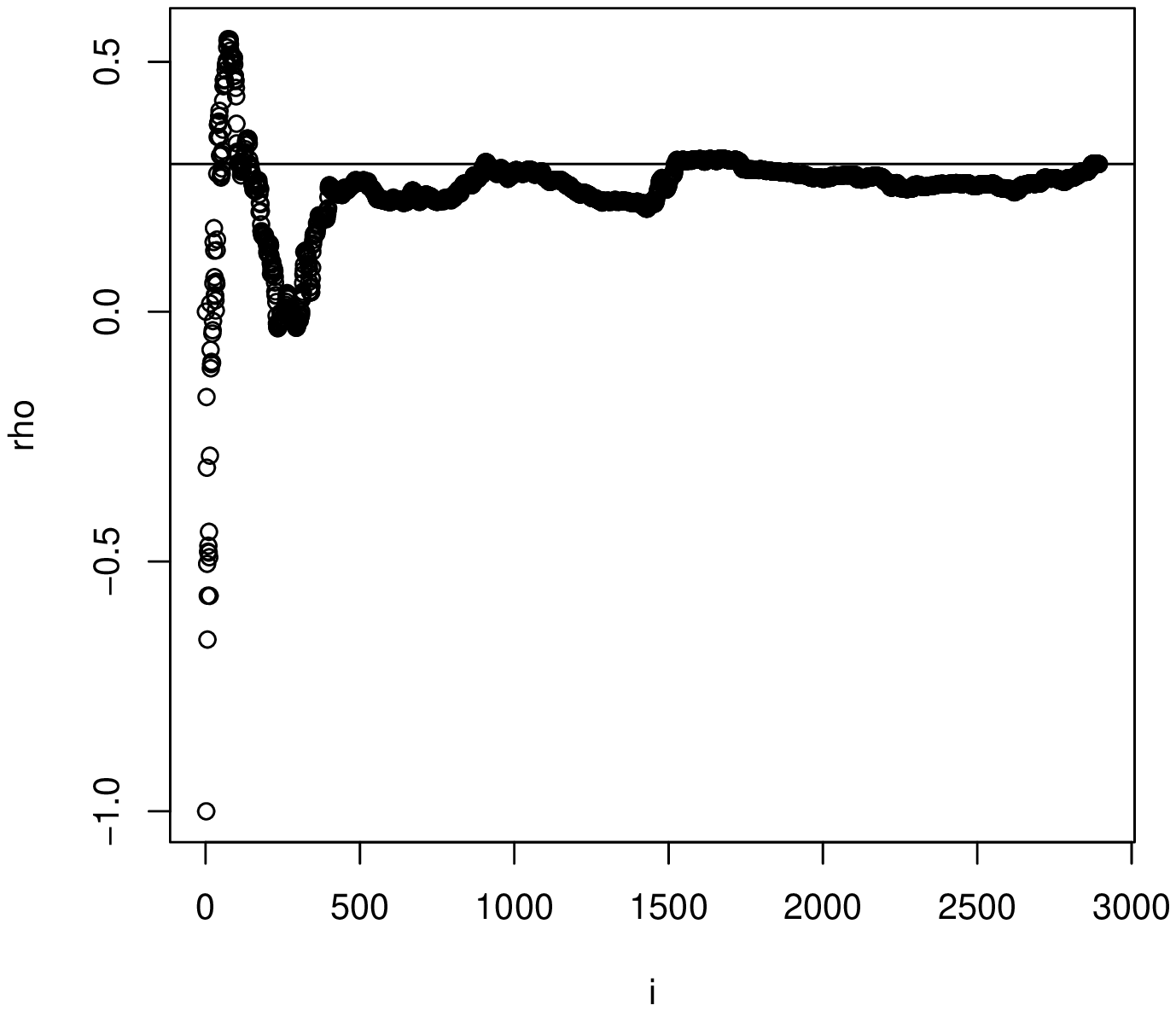}
\caption{Wave and surge heights. Dotted line represents the sample
correlations, and solid line represents the correlation estimate
$\hat{\rho}=0.2955482$ with $\hat{m}=5.614759$.}
\label{fig5}
\end{figure}

First, we calculate the $i$th sample correlation for each couple of
the mentioned data sets by using
$\{(X_{n1},Y_{n1}),\cdots,(X_{ni},Y_{ni})\}$. Figures
\ref{fig2}-\ref{fig5} show respectively that each tends to constant
ultimately. Now we estimate the correlation $\rho=1-m(i/n)/\log(n)$
by assuming that $m(s)$ is a constant, which also are illustrated by
Figures \ref{fig2}-\ref{fig5}, respectively. The constancy of $m(s)$
shows that observations are identically distributed.

\section{Proofs}
\label{sec4}

The aim of this section is to prove our main results. In the sequel,
let $F_{i}(x,y)$ denote the distribution function of
$(X_{ni},Y_{ni})$, $1\leq i\leq n$; and let $u_{n}(x)=b_{n}+x/b_{n}$
for notational simplicity.

\noindent {\bf Proof of Theorem \ref{th1}.} We only consider the
case (iii) here, since the other two cases can be derived by
Slepian's Lemma and the result of case (iii).

It follows from \eqref{eq1.3} that
\begin{equation}\label{addeq3.1}
b_{n}=(2\log n)^{\frac{1}{2}} - \frac{\log \log n + \log
4\pi}{2({2}\log n)^{\frac{1}{2}}} + o\left( \frac{1}{(\log
n)^{\frac{1}{2}}} \right),
\end{equation}
which implies that $b_{n}^{2}\sim 2\log n$ as $n\to \infty$.
Combining with \eqref{eq2.1}, we have
\begin{eqnarray}\label{eq3.1}
& & \frac{u_{n}(x)-\rho_{ni}u_{n}(z)}{\sqrt{1-\rho_{ni}^{2}}} \nonumber \\
&=& \sqrt{\frac{1-\rho_{ni}}{1+\rho_{ni}}}b_{n} + \frac{x-z}{b_{n}\sqrt{1-\rho_{ni}^{2}}}
+ \sqrt{\frac{1-\rho_{ni}}{1+\rho_{ni}}}\frac{z}{b_{n}} \nonumber\\
&=& \sqrt{\frac{m(\frac{i}{n})}{(\log n)\left( 2- \frac{m(\frac{i}{n})}{\log n} \right)}}
(2\log n)^{\frac{1}{2}}\left( 1-\frac{\log \log n + \log 4\pi}{4\log n}
+ o\left( \frac{1}{\log n} \right) \right) \nonumber \\
& & + \frac{x-z}{(2\log n)^{\frac{1}{2}}\left( 1-\frac{\log \log n + \log 4\pi}{4\log n}
+ o\left( \frac{1}{\log n} \right) \right)\sqrt{\frac{m(\frac{i}{n})}{(\log n)}\left( 2- \frac{m(\frac{i}{n})}{\log n} \right)}} \nonumber \\
& & + \sqrt{\frac{m(\frac{i}{n})}{(\log n)\left( 2- \frac{m(\frac{i}{n})}{\log n} \right)}}
\frac{z}{(2\log n)^{\frac{1}{2}}\left(1-\frac{\log \log n + \log 4\pi}{4\log n}+ o\left( \frac{1}{\log n} \right)\right)} \nonumber \\
&=& \sqrt{m\left(\frac{i}{n}\right)} + \frac{x-z}{2\sqrt{m\left(\frac{i}{n}\right)}}
- \frac{\log \log n}{4\log n}\sqrt{m\left( \frac{i}{n} \right)}
+ \frac{\log \log n}{8\log n}\frac{x-z}{\sqrt{m\left( \frac{i}{n} \right)}}
+ O\left( \frac{1}{\log n} \right) + O\left( \frac{z}{\log n} \right)
\end{eqnarray}
for large $n$.

By using the inequality $\left| \Phi(x)-\Phi(y) \right|\leq |x-y|$
for any $x,y \in \R$, we have
\begin{eqnarray*}
\left| \Phi\left( \frac{u_{n}(x)-\rho_{ni}u_{n}(z)}{\sqrt{1-\rho_{ni}^{2}}} \right)
- \Phi\left( \sqrt{m\left( \frac{i}{n} \right)
+ \frac{x-z}{2\sqrt{m\left( \frac{i}{n} \right)}}} \right) \right|
\leq (1+|z|)O\left( \frac{\log \log n}{\log n} \right)
+ (1+|z|)O\left( \frac{1}{\log n} \right)
\end{eqnarray*}
for large $n$ and any $z\in \R$, which implies that
\begin{eqnarray}\label{eq3.2}
& & \frac{1}{n} \sum_{i=1}^{n} \int_{y}^{\infty} \Phi\left( \frac{u_{n}(x)-\rho_{ni}u_{n}(z)}{\sqrt{1-\rho_{ni}^{2}}} \right)
\exp \left( -z-\frac{z^{2}}{2b_{n}^{2}} \right)dz \nonumber \\
&=& \frac{1}{n}\sum_{i=1}^{n}\int_{y}^{\infty} \Phi\left( \sqrt{m\left( \frac{i}{n} \right)}
+ \frac{x-z}{2\sqrt{m\left( \frac{i}{n} \right)}} \right) e^{-z}dz(1+o(1)) \nonumber \\
&\to& \int_{0}^{1} \int_{y}^{\infty} \Phi\left( \sqrt{m(t)}
+ \frac{x-z}{2\sqrt{m(t)}} \right){e^{-z}} dz dt \nonumber \\
&=& -e^{-x} + e^{-y}\int_{0}^{1} \Phi\left( \sqrt{m(t)} + \frac{x-y}{2\sqrt{m(t)}} \right)dt + e^{-x}\int_{0}^{1} \Phi\left( \sqrt{m(t)} + \frac{y-x}{2\sqrt{m(t)}} \right)dt
\end{eqnarray}
as $n\to \infty$.

Note that by Castro (1987) and \eqref{eq1.3},
\begin{eqnarray}\label{eq3.3}
n^{-1}=1-\Phi(b_{n}) = \frac{\varphi(b_{n})}{b_{n}}\left( 1-b_{n}^{-2}+O(b_{n}^{-4})¡¡\right)
\end{eqnarray}
for large $n$,  and Nair (1981) showed that
\begin{eqnarray}\label{eq3.4}
\lim_{n\to \infty}b_{n}^{2}\left( -n(1-\Phi(u_{n}(x))) +e^{-x} \right)
= \frac{x^{2}+2x}{2}e^{-x}.
\end{eqnarray}
Combining with \eqref{eq3.2}, we have
\begin{eqnarray}\label{eq3.5}
& & -\sum_{i=1}^{n} \left( 1-F_{i}(u_{n}(x), u_{n}(y)) \right) \nonumber \\
&=& -n\left( 1-\Phi(u_{n}(x)) \right)
- n^{-1}\left( 1 + b_{n}^{-2} + O(b_{n}^{-4}) \right)\sum_{i=1}^{n} \int_{y}^{
\infty} \Phi\left( \frac{u_{n}(x)-\rho_{ni}u_{n}(s)}{\sqrt{1-\rho_{ni}^{2}}}
 \right){\exp \left( -s - \frac{s^{2}}{2b_{n}^{2}} \right)}ds \nonumber\\
&\to& -e^{-x} - \int_{0}^{1} \int_{y}^{\infty}
\Phi\left( \sqrt{m(t)} + \frac{x-s}{2\sqrt{m(t)}} \right) e^{-s}ds dt \nonumber \\
&=& -e^{-y}\int_{0}^{1} \Phi \left( \sqrt{m(t)} + \frac{x-y}{2\sqrt{m(t)}} \right) dt - e^{-x}\int_{0}^{1} \Phi \left( \sqrt{m(t)} + \frac{y-x}{2\sqrt{m(t)}}  \right) dt
\end{eqnarray}
as $n\to \infty$, which implies the desired result.

The proof is complete.
\qed

\noindent {\bf Proof of Theorem \ref{th2}.}
By \eqref{eq3.1} we can get
\begin{eqnarray}\label{eq3.6}
& & -\frac{1}{n}\sum_{i=1}^{n} \int_{y}^{\infty} \varphi\left( \sqrt{m\left( \frac{i}{n} \right)} + \frac{x-z}{2\sqrt{m\left( \frac{i}{n} \right)}} \right)
\left( \frac{ u_{n}(x)-\rho_{ni}u_{n}(z) }{\sqrt{1-\rho_{ni}^{2}}} - \sqrt{m\left( \frac{i}{n} \right)} - \frac{x-z}{2\sqrt{m\left( \frac{i}{n} \right)}}\right) e^{-z}dz \nonumber \\
&=& \frac{\log \log n}{{4}\log n}\frac{1}{n} \sum_{i=1}^{n}
\int_{y}^{\infty} \varphi\left( \sqrt{m\left( \frac{i}{n} \right)} +
\frac{x-z}{2\sqrt{m\left( \frac{i}{n} \right)}}  \right)
\left( \sqrt{m\left( \frac{i}{n} \right)} - \frac{x-z}{2\sqrt{m\left( \frac{i}{n} \right)}} \right) e^{-z} dz + O\left( \frac{1}{\log n} \right) \nonumber \\
&\sim& \frac{\log \log n}{4\log n} \int_{0}^{1} \int_{y}^{\infty}
\varphi\left( \sqrt{m(t)} + \frac{x-z}{2\sqrt{m(t)}} \right)
\left( \sqrt{m(t)} - \frac{x-z}{2\sqrt{m(t)}} \right) e^{-z}dz dt \nonumber\\
&=& \frac{\log \log n}{2\log n}e^{-x}\int_{0}^{1} \sqrt{m(t)}\varphi\left( \sqrt{m(t)} + \frac{y-x}{2\sqrt{m(t)}} \right) dt
\end{eqnarray}
as $n\to \infty$.

By Taylor expansion with Lagrange reminder term, we have
\begin{eqnarray*}
& & \Phi\left( \frac{u_{n}(x)-\rho_{ni}u_{n}(z)}{\sqrt{1-\rho_{ni}^{2}}}  \right) \\
&=& \Phi\left( \sqrt{m\left( \frac{i}{n} \right)} + \frac{x-z}{2\sqrt{m\left( \frac{i}{n} \right)}} \right) + \varphi\left( \sqrt{m\left( \frac{i}{n} \right)}
+ \frac{x-z}{2\sqrt{m\left( \frac{i}{n} \right) }} \right)
\left( \frac{u_{n}(x)- \rho_{ni}u_{n}(z)}{2\sqrt{m\left( \frac{i}{n} \right)}}
- \sqrt{m\left( \frac{i}{n} \right)} - \frac{x-z}{2\sqrt{m\left( \frac{i}{n} \right)}}  \right) \\
& & + \frac{1}{2}\theta \varphi(\theta)\left( \frac{u_{n}(x)-\rho_{ni}u_{n}(z)}{\sqrt{1-\rho_{ni}^{2}}} - \sqrt{m\left( \frac{i}{n} \right)} - \frac{x-z}{2\sqrt{m\left( \frac{i}{n} \right)}}  \right)^{2},
\end{eqnarray*}
where \[\min\left(
\frac{u_{n}(x)-\rho_{ni}u_{n}(z)}{\sqrt{1-\rho_{ni}^{2}}},
\sqrt{m\left( \frac{i}{n} \right)} + \frac{x-z}{2\sqrt{m\left(
\frac{i}{n} \right)}} \right) < \theta < \max\left(
\frac{u_{n}(x)-\rho_{ni}u_{n}(z)}{\sqrt{1-\rho_{ni}^{2}}},
\sqrt{m\left( \frac{i}{n} \right)} + \frac{x-z}{2\sqrt{m\left(
\frac{i}{n} \right)}}  \right).\]

Combining with \eqref{eq3.6} we have
\begin{eqnarray}\label{eq3.7}
& & -\frac{1}{n}\sum_{i=1}^{n} \int_{y}^{\infty}
\left( \Phi\left( \frac{u_{n}(x)-\rho_{ni}u_{n}(z)}{\sqrt{1-\rho_{ni}^{2}}} \right)
- \Phi\left( \sqrt{m\left( \frac{i}{n} \right)} + \frac{x-z}{2\sqrt{m\left( \frac{i}{n} \right)}}  \right) \right) e^{-z} dz \nonumber \\
&=& -\frac{1}{n}\sum_{i=1}^{n}\int_{y}^{\infty} \varphi\left( \sqrt{m\left( \frac{i}{n} \right)} + \frac{x-z}{2\sqrt{m\left( \frac{i}{n} \right)}} \right)
\left( \frac{ u_{n}(x)-\rho_{ni}u_{n}(z) }{\sqrt{1-\rho_{ni}^{2}}} - \sqrt{m\left( \frac{i}{n} \right)} - \frac{x-z}{2\sqrt{m\left( \frac{i}{n} \right)}}\right) e^{-z}dz \nonumber \\
& & -\frac{1}{2n}\sum_{i=1}^{n} \int_{y}^{\infty} \theta \varphi(\theta)\left( \frac{u_{n}(x)-\rho_{ni}u_{n}(z)}{\sqrt{1-\rho_{ni}^{2}}} - \sqrt{m\left( \frac{i}{n} \right)} - \frac{x-z}{2\sqrt{m\left( \frac{i}{n} \right)}}  \right)^{2}e^{-z}dz \nonumber \\
&\sim& \frac{\log \log n}{2\log n}e^{-x}\int_{0}^{1} \sqrt{m(t)}\varphi\left( \sqrt{m(t)} + \frac{y-x}{2\sqrt{m(t)}} \right) dt
\end{eqnarray}
as $n\to \infty$ since
\[
\frac{1}{n}\sum_{i=1}^{n} \int_{y}^{\infty} \theta \varphi(\theta)\left( \frac{u_{n}(x)-\rho_{ni}u_{n}(z)}{\sqrt{1-\rho_{ni}^{2}}} - \sqrt{m\left( \frac{i}{n} \right)} - \frac{x-z}{2\sqrt{m\left( \frac{i}{n} \right)}}  \right)^{2}e^{-z}dz
=O\left( \left( \frac{\log \log n}{\log n} \right)^{2} \right).
\]

Now, we first assert that
\begin{eqnarray}\label{eq3.8}
\frac{1}{n}\sum_{i=1}^{n} \int_{y}^{\infty} \Phi\left( \sqrt{m\left( \frac{i}{n} \right)} + \frac{x-z}{2\sqrt{m\left( \frac{i}{n} \right)}}\right)e^{-z}dz
- \int_{0}^{1} \int_{y}^{\infty} \Phi\left( \sqrt{m(t)} + \frac{x-z}{2\sqrt{m(t)}} \right)e^{-z}dz dt= O\left( \frac{1}{n} \right)
\end{eqnarray}
holds for any $x,y \in \R$. Combining with \eqref{eq3.7}, we can get
\begin{eqnarray}\label{eq3.9}
& & -\frac{1}{n}\sum_{i=1}^{n} \int_{y}^{\infty} \Phi\left( \frac{u_{n}(x) - \rho_{ni}u_{n}(z)}{\sqrt{1-\rho_{ni}^{2}}} \right)e^{-z}dz
+ \int_{0}^{1} \int_{y}^{\infty} \Phi\left( \sqrt{m(t)} + \frac{x-z}{2\sqrt{m(t)}} \right)e^{-z}dz dt \nonumber \\
&\sim& \frac{\log \log n}{2\log n}e^{-x}\int_{0}^{1} \sqrt{m(t)}\varphi\left( \sqrt{m(t)} + \frac{y-x}{2\sqrt{m(t)}} \right) dt
\end{eqnarray}
as $n\to \infty$.

From \eqref{eq3.3}, \eqref{eq3.4} and \eqref{eq3.9}, it follows that
\begin{eqnarray*}
& & -\sum_{i=1}^{n}\left( 1-F_{i}(u_{n}(x), u_{n}(y)) \right) + e^{-x}
+ \int_{0}^{1} \int_{y}^{\infty} \Phi\left( \sqrt{m(t)} + \frac{x-z}{2\sqrt{m(t)}} \right)e^{-z}dz dt \\
&=& -n(1-\Phi(u_{n}(x))) + e^{-x}
- n^{-1}\left( 1+b_{n}^{-2} + O\left(b_{n}^{-4}\right) \right)
\sum_{i=1}^{n} \int_{y}^{\infty} \Phi\left( \frac{u_{n}(x) - \rho_{ni}u_{n}(z)}{\sqrt{1-\rho_{ni}^{2}}} \right)e^{-z}\exp\left( -\frac{z^{2}}{2b_{n}^{2}} \right)dz\\
& & + \int_{0}^{1} \int_{y}^{\infty} \Phi\left( \sqrt{m(t)} + \frac{x-z}{2\sqrt{m(t)}} \right)e^{-z}dz dt \\
&=& -n(1-\Phi(u_{n}(x))) + e^{-x}
-\frac{1}{n}\sum_{i=1}^{n} \int_{y}^{\infty} \Phi\left( \frac{u_{n}(x) - \rho_{ni}u_{n}(z)}{\sqrt{1-\rho_{ni}^{2}}} \right)e^{-z}dz
+ \int_{0}^{1} \int_{y}^{\infty} \Phi\left( \sqrt{m(t)} + \frac{x-z}{2\sqrt{m(t)}} \right)e^{-z}dz dt \\
& & - \frac{1}{n b_{n}^{2}}\sum_{i=1}^{n} \int_{y}^{\infty} \Phi\left( \frac{u_{n}(x) - \rho_{ni}u_{n}(z)}{\sqrt{1-\rho_{ni}^{2}}} \right)e^{-z}\left( 1- \frac{z^{2}}{2} \right)dz + O\left( b_{n}^{-4} \right) \\
&\sim& \frac{\log \log n}{2\log n}e^{-x}\int_{0}^{1} \sqrt{m(t)}\varphi\left( \sqrt{m(t)} + \frac{y-x}{2\sqrt{m(t)}} \right) dt
\end{eqnarray*}
as $n\to \infty$, which implies that
\begin{eqnarray*}
& & \P(M_{n1}\leq u_{n}(x), M_{n2}\leq u_{n}(y)) - H(x,y) \\
&=& H(x,y)\left( \exp\left( \sum_{i=1}^{n} \log F_{i}(u_{n}(x),u_{n}(y))
+ e^{-x}
+ \int_{0}^{1} \int_{y}^{\infty} \Phi\left( \sqrt{m(t)} + \frac{x-z}{2\sqrt{m(t)}} \right)e^{-z}dz dt  \right) -1 \right) \\
&=& H(x,y)\left( -\sum_{i=1}^{n}\left( 1-F_{i}(u_{n}(x), u_{n}(y)) \right) + e^{-x}
+ \int_{0}^{1} \int_{y}^{\infty} \Phi\left( \sqrt{m(t)} + \frac{x-z}{2\sqrt{m(t)}} \right)e^{-z}dz dt \right.\\
& & \left. - \frac{1}{2}\sum_{i=1}^{n}\left( 1-F_{i}(u_{n}(x), u_{n}(y)) \right)^{2}(1+o(1)) \right)(1+o(1)) \\
&\sim&  \frac{\log \log n}{2\log n}\left( \int_{0}^{1} \sqrt{m(t)}\varphi\left( \sqrt{m(t)} + \frac{y-x}{2\sqrt{m(t)}} \right) dt \right) e^{-x} H(x,y)
\end{eqnarray*}
as $n\to \infty$.

The remainder is to show that \eqref{eq3.8} holds for any fixed $x,y
\in \R$. Without loss of generality, we assume that $m(t)$ is
increasing.

If $x \leq y$, note that $\int_{y}^{\infty} \Phi\left( \sqrt{m(t)} + \frac{x-z}{2\sqrt{m(t)}}  \right)e^{-z}dz$ is increasing about $t$, so we have
\begin{eqnarray*}
& & \frac{1}{n} \sum_{i=1}^{n} \int_{y}^{\infty} \Phi\left( \sqrt{m\left( \frac{i}{n} \right) }+ \frac{x-z}{2\sqrt{m\left( \frac{i}{n} \right)}} \right)e^{-z}dz \\
&=& \sum_{i=1}^{n} \int_{\frac{i-1}{n}}^{\frac{i}{n}} \int_{y}^{\infty}
\Phi \left( \sqrt{m\left( \frac{i}{n} \right) }+ \frac{x-z}{2\sqrt{m\left( \frac{i}{n} \right)}} \right)e^{-z}dz dt\\
&>& \sum_{i=1}^{n} \int_{\frac{i-1}{n}}^{\frac{i}{n}} \int_{y}^{\infty}
\Phi \left( \sqrt{m(t)} + \frac{x-z}{2\sqrt{m(t)}} \right)e^{-z}dz dt \\
&=& \int_{0}^{1} \int_{y}^{\infty} \Phi\left( \sqrt{m(t)} + \frac{x-z}{2\sqrt{m(t)}} \right)e^{-z}dz dt
\end{eqnarray*}
and
\begin{eqnarray*}
& & \frac{1}{n} \sum_{i=1}^{n} \int_{y}^{\infty} \Phi\left( \sqrt{m\left( \frac{i}{n} \right) }+ \frac{x-z}{2\sqrt{m\left( \frac{i}{n} \right)}} \right)e^{-z}dz \\
&=& \sum_{i=1}^{n} \int_{\frac{i}{n}}^{\frac{i+1}{n}} \int_{y}^{\infty}
\Phi \left( \sqrt{m\left( \frac{i}{n} \right) }+ \frac{x-z}{2\sqrt{m\left( \frac{i}{n} \right)}} \right)e^{-z}dz dt\\
&<& \int_{0}^{1+\frac{1}{n}} \int_{y}^{\infty} \Phi\left( \sqrt{m(t)} + \frac{x-z}{2\sqrt{m(t)}} \right) e^{-z}dz dt \\
&=& \int_{0}^{1} \int_{y}^{\infty} \Phi\left( \sqrt{m(t)} + \frac{x-z}{2\sqrt{m(t)}} \right)e^{-z}dz dt + O\left( \frac{1}{n} \right),
\end{eqnarray*}
which implies that \eqref{eq3.8} holds for $x\leq y$.

To verify \eqref{eq3.8} holding for $x>y$, we just need to prove that
\begin{eqnarray}\label{eq3.10}
\frac{1}{n} \sum_{i=1}^{n} \int_{x}^{y} \Phi\left( \sqrt{m\left(
\frac{i}{n} \right) }+ \frac{x-z}{2\sqrt{m\left( \frac{i}{n}
\right)}} \right)e^{-z}dz - \int_{0}^{1} \int_{x}^{y} \Phi\left(
\sqrt{m(t)} + \frac{x-z}{2\sqrt{m(t)}} \right)e^{-z}dz dt= O\left(
\frac{1}{n} \right),
\end{eqnarray}
which will be proved in return by the following three cases: (i)
$y\leq x-2m(1)$; (ii) $x-2m(1)<y<x-2m(0)$, and (iii) $x-2m(0){\leq}
y<x$. In fact, the arguments of (i) and (iii) are similar. The rest
is to focus on (i) and (ii).

For  case (i), i.e. $y\leq x-2m(1)$, it is known that $y\leq x-2m(t)
\leq x$ for any $t \in [0,1]$. Hence,
\begin{eqnarray}\label{eq3.11}
& & \frac{1}{n} \sum_{i=1}^{n} \int_{y}^{x-2m\left( \frac{i}{n} \right)} \Phi\left( \sqrt{m\left( \frac{i}{n} \right) }+ \frac{x-z}{2\sqrt{m\left( \frac{i}{n} \right)}} \right)e^{-z}dz  \nonumber \\
&=& \sum_{i=1}^{n} \int_{\frac{i-1}{n}}^{\frac{i}{n}} \int_{y}^{x-2m\left( \frac{i}{n} \right)} \Phi\left( \sqrt{m\left( \frac{i}{n} \right) }+ \frac{x-z}{2\sqrt{m\left( \frac{i}{n} \right)}} \right)e^{-z}dz dt \nonumber \\
&<& \sum_{i=1}^{n}\int_{\frac{i-1}{n}}^{\frac{i}{n}} \int_{y}^{x-2m\left( \frac{i}{n} \right)} \Phi\left( \sqrt{m\left( t \right) }+ \frac{x-z}{2\sqrt{m\left( t \right)}} \right)e^{-z}dz dt \nonumber\\
&<& \sum_{i=1}^{n}\int_{\frac{i-1}{n}}^{\frac{i}{n}} \int_{y}^{x-2m\left( t \right)} \Phi\left( \sqrt{m\left( t \right) }+ \frac{x-z}{2\sqrt{m\left( t \right)}} \right)e^{-z}dz dt \nonumber\\
&=& \int_{0}^{1} \int_{y}^{x-2m\left( t \right)} \Phi\left( \sqrt{m\left( t \right) }+ \frac{x-z}{2\sqrt{m\left( t \right)}} \right)e^{-z}dz dt
\end{eqnarray}
and
\begin{eqnarray}\label{eq3.12}
& & \frac{1}{n} \sum_{i=1}^{n} \int_{y}^{x-2m\left( \frac{i}{n} \right)} \Phi\left( \sqrt{m\left( \frac{i}{n} \right) }+ \frac{x-z}{2\sqrt{m\left( \frac{i}{n} \right)}} \right)e^{-z}dz  \nonumber \\
&=& \sum_{i=1}^{n}\int_{\frac{i}{n}}^{\frac{i+1}{n}} \int_{y}^{x-2m\left( \frac{i}{n} \right)} \Phi\left( \sqrt{m\left(\frac{i}{n} \right) }+ \frac{x-z}{2\sqrt{m\left( \frac{i}{n} \right)}} \right)e^{-z}dz dt \nonumber \\
&>& \sum_{i=1}^{n}\int_{\frac{i}{n}}^{\frac{i+1}{n}} \int_{y}^{x-2m\left( t \right)} \Phi\left( \sqrt{m\left(\frac{i}{n} \right) }+ \frac{x-z}{2\sqrt{m\left( \frac{i}{n} \right)}} \right)e^{-z}dz dt \nonumber \\
&>& \sum_{i=1}^{n}\int_{\frac{i}{n}}^{\frac{i+1}{n}} \int_{y}^{x-2m\left( t \right)} \Phi\left( \sqrt{m\left( t \right) }+ \frac{x-z}{2\sqrt{m\left( t \right)}} \right)e^{-z}dz dt \nonumber \\
&=& \int_{0}^{1} \int_{y}^{x-2m\left( t \right)} \Phi\left( \sqrt{m\left( t \right) }+ \frac{x-z}{2\sqrt{m\left( t \right)}} \right)e^{-z}dz dt + O\left( \frac{1}{n} \right).
\end{eqnarray}

Similarly,
\begin{equation}\label{eq3.13}
\frac{1}{n} \sum_{i=1}^{n} \int_{x-2m\left( \frac{i}{n} \right)}^{x}
\Phi\left( \sqrt{m\left( \frac{i}{n} \right) }+
\frac{x-z}{2\sqrt{m\left( \frac{i}{n} \right)}} \right)e^{-z}dz \le
\int_{0}^{1} \int_{x-2m\left( t \right)}^{x} \Phi\left(
\sqrt{m\left( t \right)} + \frac{x-z}{2\sqrt{m\left( t \right)}}
\right)e^{-z}dz dt + O\left( \frac{1}{n} \right)
\end{equation}
and
\begin{eqnarray}\label{eq3.14}
 \frac{1}{n} \sum_{i=1}^{n} \int_{x-2m\left( \frac{i}{n} \right)}^{x} \Phi\left( \sqrt{m\left( \frac{i}{n} \right) }+ \frac{x-z}{2\sqrt{m\left( \frac{i}{n} \right)}} \right)e^{-z}dz
> \int_{0}^{1} \int_{x-2m\left( t \right)}^{x} \Phi\left( \sqrt{m\left( t \right)}
+ \frac{x-z}{2\sqrt{m\left( t \right)}} \right)e^{-z}dz dt.
\end{eqnarray}
Combining with \eqref{eq3.11}-\eqref{eq3.14}, it shows that
\eqref{eq3.10} holds for case (i).

Next we consider case (ii), i.e. $x-2m(1)<y<x-2m(0)$. Note that there
exists $x^{*}\in (0,1)$ such that $y=x-2m(x^{*})$ since $m(t)$ is
increasing and continuous. Split the following integral into two
parts:
\begin{eqnarray*}
& & \int_{0}^{1} \int_{y}^{x-2m\left( t \right)} \Phi\left( \sqrt{m\left( t \right) }
+ \frac{x-z}{2\sqrt{m\left( t \right)}} \right)e^{-z}dz dt  \\
&=& \int_{0}^{x^{*}} \int_{y}^{x-2m(t)}\Phi\left( \sqrt{m\left( t
\right) } + \frac{x-z}{2\sqrt{m\left( t \right)}} \right)e^{-z}dz dt
+ \int_{x^{*}}^{1} \int_{y}^{x-2m(t)}\Phi\left( \sqrt{m\left( t
\right) } + \frac{x-z}{2\sqrt{m\left( t \right)}} \right)e^{-z}dz
dt.
\end{eqnarray*}
By arguments similar with \eqref{eq3.11}-\eqref{eq3.14}, we can get
\begin{eqnarray*}
& & \frac{1}{n} \sum_{i=1}^{[nx^{*}]}\int_{y}^{x-2m\left( \frac{i}{n} \right)}
\Phi\left( \sqrt{m\left( \frac{i}{n} \right) }
+ \frac{x-z}{2\sqrt{m\left( \frac{i}{n} \right)}} \right)e^{-z}dz \\
&=& \int_{0}^{x^{*}} \int_{y}^{x-2m(t)}\Phi\left( \sqrt{m\left( t \right) }
+ \frac{x-z}{2\sqrt{m\left( t \right)}} \right)e^{-z}dz dt
+ O\left( \frac{1}{n} \right)
\end{eqnarray*}
and
\begin{eqnarray*}
& & \frac{1}{n} \sum_{i=[nx^{*}]+1}^{n}\int_{y}^{x-2m\left( \frac{i}{n} \right)}
\Phi\left( \sqrt{m\left( \frac{i}{n} \right) }
+ \frac{x-z}{2\sqrt{m\left( \frac{i}{n} \right)}} \right)e^{-z}dz \\
&=& \int_{x^{*}}^{1} \int_{y}^{x-2m(t)}\Phi\left( \sqrt{m\left( t \right) }
+ \frac{x-z}{2\sqrt{m\left( t \right)}} \right)e^{-z}dz dt
+ O\left( \frac{1}{n} \right).
\end{eqnarray*}
Combining above with \eqref{eq3.13}, \eqref{eq3.14}, we show that
\eqref{eq3.10} holds for case (ii).

Now, \eqref{eq3.10} is derived for any fixed $x,y\in \R$, which complete the proof.
 \qed

\noindent {\bf Proof of Theorem \ref{th3}.}
For fixed $x,y \in \R$, if $\max(x,y)<z<4\log b_{n}$ we have
\begin{eqnarray*}
\Phi\left(
\frac{u_{n}(\min(x,y))-\rho_{ni}u_{n}(z)}{\sqrt{1-\rho_{ni}^{2}}}
\right)  <\frac{\exp\left( -\frac{b_{n}^{2}(1-\rho_{ni})}{4}-
\frac{\min(x,y)}{1+\rho_{ni}} + \frac{\rho_{ni}z}{1+\rho_{ni}}
\right)}{\sqrt{2\pi}\left(
\frac{z-\min(x,y)}{b_{n}\sqrt{1-\rho_{ni}^{2}}}
 - \sqrt{\frac{1-\rho_{ni}}{1+\rho_{ni}}}b_{n} - \sqrt{\frac{1-\rho_{ni}}{1+\rho_{ni}}}\frac{z}{b_{n}} \right)}
\end{eqnarray*}
for large $n$ by using Mills' inequality.
Combining with \eqref{eq2.1}, \eqref{addeq3.1} and
$\lim_{n\to \infty}(\log n)^{4}\max_{1\leq i \leq n}m(i/n)=0$, we have
\begin{eqnarray*}
& & \int_{\max(x,y)}^{4\log b_{n}} \Phi\left( \frac{u_{n}(\min(x,y))
-\rho_{ni}u_{n}(z)}{\sqrt{1-\rho_{ni}^{2}}} \right) e^{-z} \exp\left( -\frac{z^{2}}{2b_{n}^{2}} \right) dz \\
&<& \frac{(1+\rho_{ni})\exp\left( -\frac{b_{n}^{2}(1-\rho_{ni})}{4} - \frac{\min(x,y)+\max(x,y)}{1+\rho_{ni}} \right)}
{\sqrt{2\pi}\left( \frac{\max(x,y)-\min(x,y)}{b_{n}\sqrt{1-\rho_{ni}^{2}}}
- \sqrt{\frac{1-\rho_{ni}}{1+\rho_{ni}}}b_{n}
- \sqrt{\frac{1-\rho_{ni}}{1+\rho_{ni}}}\frac{4\log b_{n}}{b_{n}} \right)} \\
&<& \frac{2\sqrt{2m\left( \frac{i}{n} \right)}\left( 1-\frac{\log
\log n + \log 4\pi}{{4}\log n}+ {o\left( \frac{1}{\log n} \right)}
\right) \exp\left( -\frac{1}{2}m\left( \frac{i}{n} \right)\left(
1-\frac{\log \log n + \log 4\pi}{{2}\log n} + o\left( \frac{1}{\log
n} \right)\right) + \frac{|x+y|}{2-\frac{m\left( i/n \right)}{\log
n}}\right)} {\sqrt{\pi}\left( \max(x,y) - \min(x,y) - 2m\left(
\frac{i}{n} \right)\left( 1- \frac{\log \log n + \log 4\pi}{{2}\log
n} + {o\left( \frac{1}{\log n} \right)} \right)
- 4\frac{m\left( \frac{i}{n} \right)}{\log n}\log b_{n}  \right)} \\
&<& 2\sqrt{2\max_{1\leq i \leq n}m\left( \frac{i}{n} \right)}\left(
1-\frac{\log \log n + \log 4\pi}{{4}\log n}
+ {o\left( \frac{1}{\log n} \right)} \right)\\
& & \times \frac{ \exp\left( -\frac{1}{2}\min_{1\leq i \leq
n}m\left( \frac{i}{n} \right)\left( 1-\frac{\log \log n + \log
4\pi}{{2} \log n} + o\left( \frac{1}{\log n} \right)\right) +
\frac{|x+y|}{2-\frac{\max_{1\leq i \leq n}m\left( i/n \right)}{\log
n}}\right)} {\sqrt{\pi}\left( \max(x,y) - \min(x,y) - 2m\left(
\frac{i}{n} \right)\left( 1- \frac{\log \log n + \log 4\pi}{{2} \log
n} + {o\left( \frac{1}{\log n} \right)} \right)
- 4\frac{\max_{1\leq i \leq n}m\left( i/n \right)}{\log n}\log b_{n}  \right)}\\
&=& O(b_{n}^{-4})
\end{eqnarray*}
for any $1\leq i \leq n$.

Noting that
\begin{eqnarray*}
\int_{4\log b_{n}}^{\infty} \Phi\left( \frac{u_{n}(\min(x,y))- \rho_{ni}u_{n}(z)}{\sqrt{1-\rho_{ni}^{2}}} \right)e^{-z}\exp\left( -\frac{z^{2}}{2b_{n}^{2}} \right)dz = O(b_{n}^{-4})
\end{eqnarray*}
for $1\leq i \leq n$, we have
\begin{equation*}
n^{-1}\sum_{i=1}^{n}\int_{\max(x,y)}^{\infty} \Phi\left( \frac{u_{n}(\min(x,y))- \rho_{ni}u_{n}(z)}{\sqrt{1-\rho_{ni}^{2}}} \right)e^{-z}\exp\left( -\frac{z^{2}}{2b_{n}^{2}} \right)dz = O(b_{n}^{-4})
\end{equation*}
for large $n$. Hence combining above with \eqref{eq3.4}, we can get
\begin{eqnarray*}
& & b_{n}^{2}\left(  -\sum_{i=1}^{n}\left( 1-F_{i}(u_{n}(x),u_{n}(y))
 \right) {+ e^{-\min(x,y)}} \right) \\
&=& b_{n}^{2}\left( -n(1-\Phi(\min(x,y))) + e^{-\min(x,y)}  \right) \\
& & - b_{n}^{2}n^{-1}\left( 1-b_{n}^{-2}+O(b_{n}^{-4}) \right)^{-1}
{\sum_{i=1}^{n}}\int_{\max(x,y)}^{\infty} \Phi\left(
\frac{u_{n}(\min(x,y))-\rho_{ni}u_{n}(z)} {\sqrt{1-\rho_{ni}^{2}}}
\right)e^{-z}
\exp\left(  -\frac{z^{2}}{2b_{n}^{2}}  \right)dz \\
&\to& \frac{(\min(x,y))^{2}+ 2\min(x,y)}{2}e^{-\min(x,y)}
\end{eqnarray*}
as $n\to \infty$, which implies \eqref{addeq2.3}. The proof is complete.
\qed

\noindent {\bf Proof of Theorem \ref{th4}.} By Mills' inequality we
have
\begin{eqnarray*}
1-\Phi\left(
\frac{u_{n}(y)-\rho_{ni}u_{n}(z)}{\sqrt{1-\rho_{ni}^{2}}} \right)
<\frac{\exp\left( -\frac{b_{n}^{2}(1-\rho_{ni})}{2(1+\rho_{ni})} -
\frac{y-\rho_{ni}z}{1+\rho_{ni}} - \frac{1}{2}\log
b_{n}^{2}(1-\rho_{ni}) \right)} {\sqrt{\pi}\left(
1+\frac{y-z}{b_{n}^{2}(1-\rho_{ni})} + \frac{z}{b_{n}^{2}}  \right)}
\end{eqnarray*}
for large $n$, which implies that
\begin{eqnarray*}
\int_{x}^{4\log b_{n}} \left( 1-\Phi\left(
\frac{u_{n}(y)-\rho_{ni}u_{n}(z)} {\sqrt{1-\rho_{ni}^{2}}} \right)
\right)e^{-z}\exp\left( -\frac{z^{2}}{2b_{n}^{2}}\right)dz =
O(b_{n}^{-4})
\end{eqnarray*}
for any $1\leq i \leq n$ due to \eqref{eq2.1}, \eqref{addeq3.1},
$\lim_{n\to \infty} \min_{1\leq i \leq n} m\left( i/n \right)= \infty$
and $\lim_{n\to \infty} \frac{\log \log n}{\min_{1\leq i \leq n}m(i/n)}=0$.

Combining with
\begin{eqnarray*}
\int_{4\log b_{n}}^{\infty} \left( 1-\Phi\left( \frac{u_{n}(y)-\rho_{ni}u_{n}(z)}
{\sqrt{1-\rho_{ni}^{2}}} \right) \right)e^{-z}\exp\left( -\frac{z^{2}}{2b_{n}^{2}}\right)dz = O(b_{n}^{-4}),
\end{eqnarray*}
we have
\begin{eqnarray}\label{addeq3.16}
& & \sum_{i=1}^{n}\P(X_{ni}>u_{n}(x), Y_{ni}>u_{n}(y)) \nonumber \\
&=& n^{-1}\left( 1-b_{n}^{-2} + O(b_{n}^{-4}) \right)^{-1}\sum_{i=1}^{n}
\int_{x}^{\infty} \left( 1-\Phi\left( \frac{u_{n}(y)-\rho_{ni}u_{n}(z)}
{\sqrt{1-\rho_{ni}^{2}}} \right) \right)e^{-z}\exp\left( -\frac{z^{2}}{2b_{n}^{2}}\right)dz \nonumber \\
&=& O(b_{n}^{-4}).
\end{eqnarray}
It follows from \eqref{eq3.4} and \eqref{addeq3.16} that
\begin{eqnarray*}
& & b_{n}^{2}\left( -\sum_{i=1}^{n} (1-F_{i}(u_{n}(x),u_{n}(y))) +e^{-x} +e^{-y} \right)\\
&=& b_{n}^{2}\left( -n(1-\Phi(u_{n}(x))) + e^{-x} -n(1-\Phi(u_{n}(y))) + e^{-y}
+ \sum_{i=1}^{n}\P(X_{ni}>u_{n}(x), Y_{ni}>u_{n}(y)) \right) \\
&\to& \frac{x^{2}+2x}{{2}}e^{-x} +\frac{y^{2}+2y}{2}e^{-y}
\end{eqnarray*}
as $n\to \infty$. Hence \eqref{addeq2.4} can be derived, which complete the proof.
\qed

\noindent {\bf Proof of Theorem \ref{th5}.}
Define
\begin{eqnarray*}
Z_{i}
&=& - \frac{\rho_{ni}}
{(\log n)\left( 1 - \rho_{ni}^{2}  \right)^{2} }\left( X_{ni}^{2} + Y_{ni}^{2} \right)
+ \frac{1+\rho_{ni}^{2}}{(\log n)(1- \rho_{ni}^{2})^{2}}X_{ni}Y_{ni}
+\frac{\rho_{ni}}{(\log n)(1-\rho_{ni}^{2})} \\
&:=& Z_{i,1} + Z_{i,2} + Z_{i,3},
\end{eqnarray*}
one can check that
\begin{eqnarray*}
\begin{array}{llll}
\E Z_{i,1}^{2}& = \frac{4\left( \rho_{ni}^{4} + 2\rho_{ni}^{2}
\right)} {(\log n)^{2}\left( 1-\rho_{ni}^{2} \right)^{4}}, \quad
&\E Z_{i,2}^{2}&= \frac{(1+\rho_{ni}^{2})^{2}}{(\log n)^{2}(1-\rho_{ni}^{2})^{4}}(1+2\rho_{ni}^{2}) \\
\E Z_{i,3}^{2}& = \frac{\rho_{ni}^{2}}{(\log
n)^{2}(1-\rho_{ni}^{2})^{{2}}}, \quad &\E Z_{i,1}Z_{i,2}&= -
\frac{6\rho_{ni}^{2}(1+\rho_{ni}^{2})}
{(\log n)^{2}(1-\rho_{ni}^{2})^{4}} \\
\E Z_{i,1}Z_{i,3}&= - \frac{2\rho_{ni}^{2}}{(\log
n)^{2}(1-\rho_{ni}^{2})^{3}}, \quad &\E Z_{i,2}Z_{i,3}&=
\frac{\rho_{ni}^{2}{(1+\rho_{ni}^{2})}}{(\log
n)^{2}(1-\rho_{ni}^{2})^{3}},
\end{array}
\end{eqnarray*}
which implies that
\begin{eqnarray}\label{eq3.15}
\frac{1}{n}\sum_{i=1}^{n} \E Z_{i}^{2} &=& \frac{1}{n}
\sum_{i=1}^{n} \left( \frac{1-\frac{\alpha + \beta \left(
\frac{i}{n} \right)^{\gamma}}{\log n} + \frac{\left( \alpha + \beta
\left( \frac{i}{n} \right)^{\gamma} \right)^{2}}{2(\log n)^{2}}}
{2\left( \alpha + \beta \left( \frac{i}{n} \right)^{\gamma}
\right)^{2}
\left( 1-\frac{\alpha + \beta \left( \frac{i}{n} \right)^{\gamma}}{2\log n} \right)^{2}} \right) \nonumber \\
&\to& \int_{0}^{1} \frac{1}{2(\alpha + \beta t^{\gamma})^{2}}dt
\end{eqnarray}
as $n\to \infty$. It is easy to check that
\begin{eqnarray*}
\E\left( \frac{1}{n}\sum_{i=1}^{n} \left( Z_{i}^{2} - \E Z_{i}^{2}
\right)¡¡\right)^{2} = \frac{1}{n^{2}}\sum_{i=1}^{n} \E\left(
Z_{i}^{2} - \E Z_{i}^{2} \right)^{2} = O\left( \frac{1}{n} \right),
\end{eqnarray*}
which combining with \eqref{eq3.15} implies that
\begin{eqnarray}\label{eq3.16}
\sum_{i=1}^{n}\left( \frac{1}{\sqrt{n}} Z_{i} \right)^{2} \overset{p}{\to}
\int_{0}^{1} \frac{1}{2(\alpha + \beta t^{\gamma})^{2}}dt
\end{eqnarray}
Obviously, we have
\begin{eqnarray}\label{eq3.17}
\max_{1\leq i \leq n} \left| \frac{1}{\sqrt{n}}Z_{i} \right|
\overset{p}{\to} 0 \quad \text{and} \quad \E\left( \max_{1\leq i
\leq n} \frac{1}{n}Z_{i}^{2} \right)=o(1).
\end{eqnarray}
Hence combining with \eqref{eq3.16}, \eqref{eq3.17} we can get
\begin{equation}\label{eq3.18}
\frac{1}{\sqrt{n}} l_{n1}\left( \alpha, \beta, \gamma \right) \overset{p}{\to}
N\left( 0, \int_{0}^{1} \frac{1}{2(\alpha + \beta t^{\gamma})^{2}}dt\right)
\end{equation}
as $n\to \infty$.

Define
\begin{eqnarray*}
Z_{i}^{*}&=& - \frac{\left( \frac{i}{n} \right)^{\gamma}\rho_{ni}}
{(\log n)\left( 1 - \rho_{ni}^{2}  \right)^{2} }\left( X_{ni}^{2} + Y_{ni}^{2} \right)
+ \frac{\left( \frac{i}{n} \right)^{\gamma}(1+\rho_{ni}^{2})}{(\log n)(1- \rho_{ni}^{2})^{2}}X_{ni}Y_{ni}
+\frac{\left( \frac{i}{n} \right)^{\gamma} \rho_{ni}}{(\log n)(1-\rho_{ni}^{2})} \\
&:=& Z_{i,1}^{*} + Z_{i,2}^{*} + Z_{i,3}^{*},
\end{eqnarray*}
and
\begin{eqnarray*}
Z_{i}^{**}&=& - \frac{\left( \frac{i}{n} \right)^{\gamma}\left( \log
\frac{i}{n} \right)\rho_{ni}} {(\log n)\left( 1 - \rho_{ni}^{2}
\right)^{2} }\left( X_{ni}^{2} + Y_{ni}^{2} \right) + \frac{{\left(
\frac{i}{n} \right)^{\gamma}\left( \log \frac{i}{n}
\right)(1+\rho_{ni}^{2})}}{(\log n)(1-
\rho_{ni}^{2})^{2}}X_{ni}Y_{ni}
+\frac{\left( \frac{i}{n} \right)^{\gamma}\left( \log \frac{i}{n} \right) \rho_{ni}}{(\log n)(1-\rho_{ni}^{2})} \\
&:=& Z_{i,1}^{**} + Z_{i,2}^{**} + Z_{i,3}^{**}.
\end{eqnarray*}

Similar to the proofs of \eqref{eq3.18}, we can show that
\begin{equation}\label{eq3.19}
\left\{
\begin{array}{ll}
\lim_{n\to \infty} \frac{1}{\sqrt{n}}l_{n2}(\alpha,\beta,\gamma)
& \overset{p}{=} N\left( 0, \int_{0}^{1} \frac{t^{2\gamma}}{2\left(\alpha +\beta t^{\gamma}\right)^{2}} {dt} \right), \\
\lim_{n\to \infty} \frac{1}{\sqrt{n}}l_{n3}(\alpha,\beta,\gamma) &
\overset{p}{=} N\left( 0, \int_{0}^{1} \frac{t^{2\gamma}(\log
t)^{2}}{2\left(\alpha +\beta t^{\gamma}\right)^{2}} {dt} \right).
\end{array}
\right.
\end{equation}

By arguments similar to \eqref{eq3.15}, we have
\begin{eqnarray*}
\lim_{n\to \infty}\frac{1}{n}\sum_{i=1}^{n} \E Z_{i}Z_{i}^{*} =
\lim_{n\to \infty}\frac{1}{n}\sum_{i=1}^{n} \left( \frac{i}{n}
\right)^{\gamma}\E Z_{i}^{2} = \int_{0}^{1}
\frac{t^{\gamma}}{2(\alpha + \beta t^{\gamma})^{2}} dt,
\end{eqnarray*}
\begin{eqnarray*}
\lim_{n\to \infty} \frac{1}{n}\sum_{i=1}^{n} \E Z_{i}^{*}Z_{i}^{**}
= \lim_{n\to \infty} \frac{1}{n}\sum_{i=1}^{n} \left( \frac{i}{n}
\right)^{2\gamma}\left(\log \frac{i}{n}\right)\E Z_{i}^{2} =
\int_{0}^{1} \frac{t^{2\gamma}\log t}{2(\alpha + \beta
t^{\gamma})^{2}} dt
\end{eqnarray*}
and
\begin{eqnarray*}
\lim_{n\to \infty} \frac{1}{n}\sum_{i=1}^{n} \E Z_{i}Z_{i}^{**} =
\lim_{n\to \infty} \frac{1}{n}\sum_{i=1}^{n} \left( \frac{i}{n}
\right)^{\gamma}\left(\log \frac{i}{n}\right) \E Z_{i}^{2} =
\int_{0}^{1} \frac{t^{\gamma}\log t}{2(\alpha + \beta
t^{\gamma})^{2}} dt.
\end{eqnarray*}
Hence, by Cram\'{e}r device, we can derive that
\begin{eqnarray}\label{eq3.20}
\lim_{n\to \infty}\frac{1}{\sqrt{n}} \left( l_{n1}(\alpha, \beta, \gamma), l_{n2}(\alpha, \beta, \gamma), l_{n3}(\alpha, \beta, \gamma) \right)^{T} \overset{d}{=}
N\left( 0, \Sigma \right),
\end{eqnarray}
where $\Sigma$ is given by \eqref{eq2.5}.

It is straight forward to check that
\begin{eqnarray}\label{eq3.21}
& & \lim_{n\to \infty}\frac{1}{n} \left(
\begin{array}{lll}
\frac{\partial l_{n1}(\alpha, \beta, \gamma)}{\partial \alpha}
& \frac{\partial l_{n1}(\alpha, \beta, \gamma)}{\partial \beta}
& \frac{\partial l_{n1}(\alpha, \beta, \gamma)}{\partial \gamma} \\
\frac{\partial l_{n2}(\alpha, \beta, \gamma)}{\partial \alpha}
& \frac{\partial l_{n2}(\alpha, \beta, \gamma)}{\partial \beta}
& \frac{\partial l_{n2}(\alpha, \beta, \gamma)}{\partial \gamma} \\
\frac{\partial l_{n3}(\alpha, \beta, \gamma)}{\partial \alpha}
& \frac{\partial l_{n3}(\alpha, \beta, \gamma)}{\partial \beta}
& \frac{\partial l_{n3}(\alpha, \beta, \gamma)}{\partial \gamma} \\
\end{array}
\right) \nonumber\\
& &\overset{p}{=}
\begin{pmatrix}
\int_{0}^{1}\frac{1}{2(\alpha+\beta t^{\gamma})^{2}}dt
&\int_{0}^{1}\frac{t^{\gamma}}{2(\alpha+\beta t^{\gamma})^{2}}dt
&\int_{0}^{1}\frac{\beta t^{\gamma}\log t}{2(\alpha+\beta t^{\gamma})^{2}}dt\\
\int_{0}^{1}\frac{t^{\gamma}}{2(\alpha+\beta t^{\gamma})^{2}}dt
&\int_{0}^{1}\frac{t^{2\gamma}}{2(\alpha+\beta t^{\gamma})^{2}}dt
&\int_{0}^{1}\frac{{\beta}t^{2\gamma}\log t}{2(\alpha+\beta t^{\gamma})^{2}}dt\\
\int_{0}^{1}\frac{{t^{\gamma}\log t}}{2(\alpha+\beta
t^{\gamma})^{2}}dt &\int_{0}^{1}\frac{t^{2\gamma}\log
t}{2(\alpha+\hat{\beta} t^{\gamma})^{2}}dt &\int_{0}^{1}\frac{\beta
t^{2\gamma}(\log t)^2}{2(\alpha+\beta t^{\gamma})^{2}}dt
\end{pmatrix} \nonumber \\
& &:=
 \Delta.
\end{eqnarray}
Hence, the desired result is derived by \eqref{eq3.20}, \eqref{eq3.21} and
Taylor expansion. The proof is complete.
\qed

\noindent {\bf Proof of Theorem \ref{th6}.}
It follows from the proof of Theorem \ref{th5} with known $\gamma =1$.
\qed

\vspace{1cm}

\noindent {\bf Acknowledgements}~~ This work was supported by the
National Natural Science Foundation of China grant no.11171275, the
Natural Science Foundation Project of CQ no. cstc2012jjA00029, and
the Fundamental Research Funds for the Central
Universities(XDJK2014D020).

\end{document}